\documentclass[11pt,a4paper]{article}

\usepackage[T1]{fontenc}
\usepackage{stix2} 
\usepackage{microtype}

\usepackage[margin=1in]{geometry}
\usepackage{setspace}
\usepackage{authblk}

\usepackage{amsmath}
\usepackage{mathtools}
\usepackage{mathrsfs} 
\usepackage{amssymb}
\usepackage{amsfonts}
\usepackage{bm}   

\usepackage{tikz}
\usepackage{graphicx}
\usepackage{float}
\usepackage[dvipsnames,svgnames,x11names]{xcolor}

\usepackage{booktabs}
\usepackage{multirow}
\usepackage{tabularx}
\usepackage[labelsep=period]{caption}

\usepackage[colorlinks=true,
            linkcolor=NavyBlue,
            citecolor=ForestGreen,
            urlcolor=BrickRed]{hyperref}

\usepackage[capitalize,noabbrev]{cleveref}

\usepackage{algorithm}
\usepackage{algpseudocode}

\usepackage{listings}
\usepackage{tcolorbox}
\tcbuselibrary{skins,breakable}

\usepackage{xspace}

\usepackage[numbers,sort&compress]{natbib}
\usepackage{macros}

\title{
  Global-in-time Spectral Duhamel--Picard Method for Post-Exposure Bake Simulation and Physics-Informed Neural Operators for EUV Lithography
}

\author[]{Doyun Kim\thanks{doyun.kim@imec.be}}
\author[]{Mircea Dusa}
\author[]{Werner Gillijns}
\author[]{Philippe Leray}
\affil[]{IMEC, Kapeldreef 75, 3001 Leuven, Belgium}
\date{}

\begin{document} 

\maketitle

\begin{abstract}
We present a global-in-time spectral Duhamel--Picard method with polynomial expansions for post-exposure bake (PEB) simulation.  
The coupled acid--quencher reaction--diffusion equations are reformulated as nonlinear Volterra integral equations using the Duhamel principle.  
Full trajectories are represented by finite polynomial expansions, with temporal coefficient fields determined over the bake interval by Anderson-accelerated Picard iteration, thereby avoiding sequential time marching.
In 3-D PEB simulations, we employ Chebyshev and Legendre temporal bases while evaluating the Duhamel integrals using the corresponding Gaussian quadrature rules.
We demonstrate that the proposed method simulates a large domain of $1200\times1200\times100$ voxels on the order of one minute.
The representation is further incorporated into a physics-informed neural operator (PINO) with a mixed-transform variant of a factorized Fourier neural operator, which is trained without precomputed solution trajectories.
Together, the integral formulation and its PINO extension enable accelerated, time-marching-free prediction of concentration fields at arbitrary times within the bake interval.

\end{abstract}

\section{Introduction}
\label{sec:introduction}

Post-exposure bake (PEB) is a critical thermal step in extreme-ultraviolet (EUV) lithography with chemically amplified resists (CARs), during which acid diffusion, quencher neutralisation, and deprotection reactions take place, ultimately determining the printed pattern fidelity~\cite{mack2018fundamental,postEBbake_review}.
Accurate simulation of the PEB reaction--diffusion system is therefore fundamental to quantitatively characterising resist behaviour and optimising lithographic processes.
Such simulations are particularly challenging in large-area optical proximity correction (OPC), where demanding computational requirements have motivated the widespread use of calibrated compact resist models. 
Typically formulated on predominantly two-dimensional layout domains, these models approximate the coupled effects of aerial-image formation, diffusion, reaction kinetics, resist shrinkage, and development through calibrated model terms, including convolutional kernels and nonlinear response functions~\cite{kimura2020compact,park2025morcompact,granik2020shrinkage}.
Consequently, rigorous three-dimensional PEB reaction--diffusion simulations have not been routinely integrated into iterative OPC loops, despite their value for model calibration and validation~\cite{lin2013peb}.

Numerically, PEB simulation is commonly formulated as a coupled partial differential equation (PDE) system and solved by explicit time marching, for which the stable time step generally scales as $\Delta t_{\max}\propto\Delta x^2/D$.
Consequently, fine spatial grids or large diffusivities require a large number of sequential time steps over the full bake interval, each involving the evaluation of the nonlinear reaction terms and diffusion operator throughout the three-dimensional domain.
This computational burden is especially pronounced in model calibration and high-throughput process optimisation, even when only the final-state concentration distributions are of practical interest.

Conventional numerical methods have primarily focused on improving the accuracy and efficiency of time marching.  
Higher order spatial discretisations include compact finite difference stencils with fourth or sixth order accuracy~\cite{fornberg1988generation}.  
Explicit Runge-Kutta time integrators also improve the accuracy of individual steps and can reduce the total step count.  

PEB-specific learning-based surrogates have also been developed to reduce the cost of conventional simulations.
DeePEB combines a Fourier neural-operator branch with a local convolutional branch and is trained in a supervised manner on PEB solutions generated by a numerical simulator~\cite{wang2022deepeb}.
It directly maps an initial photoacid distribution to the inhibitor distribution at a prescribed final bake time without explicitly marching the governing PDEs in time.
The reported model is therefore designed to predict the final PEB state rather than concentration fields at arbitrary times within the PEB interval.
Cho \emph{et al.}\ use a convolutional long short-term memory (convLSTM) network that learns a recurrent state-update rule by minimizing a PDE-based loss and applies that rule sequentially over the bake duration~\cite{cho2024rnnpeb}.
It consequently retains a learned form of time marching at inference.

In this work, we develop a global-in-time spectral method and a physics-informed neural operator (PINO) that replace sequential time marching with a global temporal representation for efficient PEB simulation.
The proposed method employs an integral formulation and high-order temporal quadrature, as in spectral deferred correction~\cite{dutt2000sdc} and numerical methods for Volterra integral equations~\cite{linz1985volterra}.
To extend this integral--quadrature approach to the coupled PEB reaction--diffusion system, we recast the governing equations in a Duhamel form that incorporates linear diffusion through the heat semigroup. 
The resulting nonlinear acid and quencher trajectories are then represented by global polynomial expansions, whose temporal coefficient fields are determined using a combination of Anderson-accelerated Picard iteration and Gaussian quadrature.
Thus, the temporal solution is obtained globally over the bake interval rather than through a sequence of time steps. 
Both Chebyshev and Legendre polynomial bases are examined, together with the corresponding Gauss--Chebyshev and Gauss--Legendre quadrature rules.
We further use the global-in-time spectral Duhamel formulation to construct the physics-informed loss for a PINO with a mixed-transform variant of the factorized Fourier neural operator (f-FNO) backbone. 
Precomputed simulation trajectories or experimental measurements are not used for training, and the PINO predicts the temporal coefficient fields in a single forward pass at inference.
The resulting concentration distributions can then be evaluated at arbitrary times within the bake interval without sequential time marching, which is useful for model calibration and process optimization.
Numerical results demonstrate substantial acceleration of the global-in-time spectral solver over sequential time marching, while the PINO reliably predicts the concentration trajectories throughout the bake interval with further reduction in computational time.

\section{Governing Equations and Solution Method}
\label{sec:governing_equations}

\subsection{Duhamel Formulation for PEB Reaction-Diffusion System}
\label{sec:Duhamel_spectral}
The PEB process evolves four concentration fields---acid $\acid$, quencher $\quench$, protected group $\prot$, and deprotected group $\deprot$---governed by the reaction--diffusion system
\begin{align}
  \pder{\prot}{t}
    &= -k_1\prot^{p}\acid^{q} - k_2\prot,
    \label{eq:P}\\
  \pder{\deprot}{t}
    &= k_1\prot^{p}\acid^{q} + k_2\prot + \Ddep\lapl\deprot,
    \label{eq:D}\\
  \pder{\acid}{t}
    &= -k_3\acid^{r} - k_4\acid\quench + \Dacid\lapl\acid,
    \label{eq:A}\\
  \pder{\quench}{t}
    &= -k_5\quench^{s} - k_4\acid\quench + \Dquench\lapl\quench.
    \label{eq:Q}
\end{align}

The four fields are not equally coupled. The $\prot$ and $\deprot$ fields depend on the coupled $\acid$--$\quench$ subsystem and do not feed back into either $\acid$ or $\quench$, allowing the system to be solved hierarchically. 
We first solve for the coupled $\acid$ and $\quench$ trajectories and then evaluate $\prot$ and $\deprot$ sequentially using semi-analytical solutions.

Identifying the nonlinear reaction sinks in \cref{eq:A,eq:Q} as inhomogeneous forcing terms, the solution decomposes by superposition into a homogeneous part, corresponding to source-free spatial diffusion of the initial distribution, and an inhomogeneous part, corresponding to the reaction sinks. 
The Duhamel formula
~\cite{evans2010pde,pazy1983semigroups} provides this decomposition
explicitly. Applying the Duhamel formula to \cref{eq:A} gives
\begin{equation}
\begin{split}
  \acid(\mathbf{x},t)
    &= \int_{\Omega}
         \Kacid^{\mathcal B}(\mathbf{x},\mathbf{x}';t)
         \Azero(\mathbf{x}')
       \intd\mathbf{x}' \\
    &\quad -
       \int_0^{t}\!\int_{\Omega}
         \Kacid^{\mathcal B}(\mathbf{x},\mathbf{x}';t-s)
         \bigl(k_3\acid^r+k_4\acid\quench\bigr)(\mathbf{x}',s)
       \intd\mathbf{x}'\intd s.
\end{split}
  \label{eq:A_semigroup}
\end{equation}
Here, $\Omega$ denotes the computational domain, $\mathcal{B}$ represents the imposed spatial boundary conditions, $t\in[0,t_{PEB}]$ is the physical evaluation time, $s$ is the time-integration variable, and $t_{\mathrm{PEB}}$ is the final post-exposure bake time.
Unlike the free-space heat kernel, $\Kacid^{\mathcal B}$ need not depend only on $\mathbf{x}-\mathbf{x}'$ when non-periodic boundary conditions are imposed in the $z$ direction.

We evaluate the spatial heat-kernel convolutions spectrally using the 2-D Fast Fourier Transform (FFT) $\mathbf{F}_{xy}$ in the periodic $x$--$y$ directions and the $z$-direction discrete transform $\mathbf{T}_{z}\in\{\text{DCT},\text{DST}\}$, namely a discrete cosine transform (DCT) for Neumann boundary conditions or a discrete sine transform (DST) for Dirichlet boundary conditions. 
Diffusion is then evaluated by pointwise multiplication in spectral space. 
Explicitly,
\begin{equation}
\begin{split}
  \acid(\mathbf{x},t)
    &= \mathbf{F}_{xy}^{-1}\mathbf{T}_{z}^{-1}\!\left[
         \hat{\mathbf{K}}_A(\mathbf{k},t)
         \hat{A}_0(\mathbf{k})
       \right](\mathbf{x}) \\[4pt]
    &\quad -
       \int_0^{t}
       \mathbf{F}_{xy}^{-1}\mathbf{T}_{z}^{-1}\!\Bigl[
         \hat{\mathbf{K}}_A(\mathbf{k},t-s)
         \mathbf{F}_{xy}\mathbf{T}_{z}\!\left[
           k_3\acid^r(\cdot,s)
           +k_4\acid(\cdot,s)\quench(\cdot,s)
         \right](\mathbf{k})
       \Bigr](\mathbf{x})\intd s.
\end{split}
  \label{eq:A_spectral}
\end{equation}
Here,
\begin{equation}
  \mathbf{k}=(k_x,k_y,k_z)
  \label{eq:spatial_mode_vector}
\end{equation}
denotes the spatial spectral-mode vector, and
\begin{equation}
  \hat{u}(\mathbf{k})
  =
  \mathbf{F}_{xy}\mathbf{T}_{z}u
  \label{eq:mixed_spatial_transform}
\end{equation}
denotes the mixed spatial transform of any concentration field
$u\in\{A,Q,P,D\}$.

The heat semigroup is diagonal in this basis:
\begin{equation}
  \hat{\mathbf{K}}_A(\mathbf{k},t)
  =
  \exp\!\left[
    -\Dacid\left(k_x^2+k_y^2+k_z^2\right)t
  \right].
  \label{eq:heat_kernel_A}
\end{equation}
Accordingly, $k_x$ and $k_y$ are Fourier wavenumbers, while in the $z$ direction $k_z=n\pi/L_z$ indexes DCT cosine modes with $n=0,\ldots,N_z-1$ or DST sine modes with $n=1,\ldots,N_z$.
This spectral implementation reduces the cost of heat-semigroup evaluations while retaining spectral accuracy for sufficiently smooth spatial fields~\cite{gottlieb1977numerical,canuto1988spectral}.
An analogous formula holds for $\quench$. In the inhomogeneous part of \cref{eq:A_spectral}, $\acid$ at any time $t$ depends on its own past trajectory and that of $\quench$, making the equation \emph{non-local in
time}. 
The $\acid$ and $\quench$ trajectories must therefore be determined in a mutually consistent manner.

Conventional explicit solvers resolve this temporal coupling by time marching, with the time-step size constrained by the spatial diffusion scales.
In \cref{sec:chebyshev_picard}, we instead represent and solve the coupled trajectories over the entire bake interval using global Chebyshev or Legendre polynomial bases, without sequential time marching.
Once $\acid(\mathbf{x},t)$ is known for all $t$, \cref{eq:P} becomes a \emph{Bernoulli} ODE in $\prot$ alone, driven by the known field $\acid^q$.
For $p=1$, \cref{eq:P} reduces to a linear first-order ODE with the
closed-form solution
\begin{equation}
  \prot(\mathbf{x},t)
  =
  \Pzero(\mathbf{x})
  \exp\!\left[
    -k_1\int_0^t\acid^q(\mathbf{x},s)\intd s
    -k_2t
  \right],
  \qquad p=1.
  \label{eq:P_linear}
\end{equation}
For $p\neq1$, the substitution $v=\prot^{1-p}$ transforms
\cref{eq:P} into a linear ODE, giving
\begin{equation}
  \prot(\mathbf{x},t)
  =
  \Biggl[
    e^{-\beta t}
    \left(
      \bigl[\Pzero(\mathbf{x})\bigr]^{1-p}
      -(1-p)k_1
      \int_0^t e^{\beta s}\acid^q(\mathbf{x},s)\intd s
    \right)
  \Biggr]^{1/(1-p)},
  \qquad p\neq1,
  \label{eq:P_preview}
\end{equation}
where $\beta=k_2(1-p)$.

With $\acid$ and $\prot$ available, the Duhamel form of \cref{eq:D} gives
\begin{equation}
\begin{split}
  \deprot(\mathbf{x},t)
    &= \mathbf{F}_{xy}^{-1}\mathbf{T}_{z}^{-1}\!\left[
         \hat{\mathbf{K}}_D(\mathbf{k},t)
         \hat{D}_0(\mathbf{k})
       \right](\mathbf{x}) \\[4pt]
    &\quad +
       \int_0^t
       \mathbf{F}_{xy}^{-1}\mathbf{T}_{z}^{-1}\!\Bigl[
         \hat{\mathbf{K}}_D(\mathbf{k},t-s)
         \mathbf{F}_{xy}\mathbf{T}_{z}\!\left[
           k_1\prot^p(\cdot,s)\acid^q(\cdot,s)
           +k_2\prot(\cdot,s)
         \right](\mathbf{k})
       \Bigr](\mathbf{x})\intd s.
\end{split}
  \label{eq:D_preview}
\end{equation}
Here, the deprotected-group heat semigroup is
\begin{equation}
  \hat{\mathbf{K}}_D(\mathbf{k},t)
  =
  \exp\!\left[
    -\Ddep\left(k_x^2+k_y^2+k_z^2\right)t
  \right].
  \label{eq:heat_kernel_D}
\end{equation}

Equations~\eqref{eq:P_linear} and \eqref{eq:P_preview} determine $\prot$
directly from the prescribed trajectory of $\acid^q$ for $p=1$ and
$p\neq1$, respectively.
The source term in
\cref{eq:D_preview} then depends only on the already-determined trajectories
of $\acid$ and $\prot$ and does not depend on $\deprot$. Consequently,
neither field requires an additional Picard iteration once the acid and
quencher trajectories are known.

\subsection{Global-in-Time Picard Iteration with Polynomials and Gaussian Quadrature}
\label{sec:chebyshev_picard}
\cref{eq:A_spectral} and its analogue for $\quench$ are coupled nonlinear
Volterra integral equations. We solve the acid and quencher trajectories
jointly by Anderson-accelerated Picard iteration
~\cite{anderson1965iterative,walker2011anderson,toth2015convergence}.
Each Picard sweep applies the Duhamel operator to the current trajectories
to form a new self-consistent estimate. 
Anderson acceleration combines recent Picard updates to accelerate the reduction of the fixed-point residual.

Following Gaussian-quadrature-based approaches for Volterra integral equations~\cite{linz1985volterra}, we evaluate the Duhamel time integral at an arbitrary time $t\in[0,t_{PEB}]$ using an $N$-point Gaussian quadrature rule.
Let $\{(\xi_i,w_i)\}_{i=1}^{N}$ denote the reference quadrature nodes and weights on $[-1,1]$. Mapping this rule to $[0,t]$ gives
\begin{equation}
  s_i(t)=\frac{t}{2}(1+\xi_i),
  \qquad
  w_i(t)=\frac{t}{2}w_i,
  \qquad
  \int_0^t f(s)\intd s
  \approx
  \sum_{i=1}^{N}w_i(t)f\!\left(s_i(t)\right).
  \label{eq:quadrature_map}
\end{equation}
Here, $t$ is an arbitrary evaluation time, and $N$ is the quadrature order. 
In this study, the reference nodes and weights $\{(\xi_i,w_i)\}_{i=1}^{N}$ are chosen from either Gauss--Chebyshev quadrature of the first kind~\cite{trefethen2000spectral} or Gauss--Legendre quadrature~\cite{abramowitz1965handbook}. 
For the Gauss--Chebyshev rule, the standard weights are adjusted to approximate the ordinary unweighted Duhamel integral.
To illustrate, \cref{tab:quadrature_nodes} compares the mapped nodes and weights of both
rules over $[0,t_{\mathrm{PEB}}]$ for $N=8$ and $t_{\mathrm{PEB}}=100\,\mathrm{s}$.

\begin{table}[h]
  \centering
  \caption{Comparison of $N=8$ Gauss--Chebyshev and Gauss--Legendre
           quadrature nodes, physical times, and scaled weights over the
           full bake interval $[0,t_{PEB}]$ for $t_{PEB}=100\,\mathrm{s}$.}
  \label{tab:quadrature_nodes}
  \begin{tabular}{crrrrrr}
    \toprule
    & \multicolumn{3}{c}{Gauss--Chebyshev}
    & \multicolumn{3}{c}{Gauss--Legendre} \\
    \cmidrule(lr){2-4}\cmidrule(lr){5-7}
    $i$ & $\xi_i$ & $s_i(t_{PEB})\;[\mathrm{s}]$ & $w_i(t_{PEB})\;[\mathrm{s}]$
        & $\xi_i$ & $s_i(t_{PEB})\;[\mathrm{s}]$ & $w_i(t_{PEB})\;[\mathrm{s}]$ \\
    \midrule
    1 & $+0.980785$ & $99.039$            & $\phantom{0}3.831$
      & $+0.960290$ & $98.014$            & $\phantom{0}5.061$ \\
    2 & $+0.831470$ & $91.574$            & $10.909$
      & $+0.796666$ & $89.833$            & $11.119$ \\
    3 & $+0.555570$ & $77.779$            & $16.326$
      & $+0.525532$ & $76.277$            & $15.685$ \\
    4 & $+0.195090$ & $59.755$            & $19.258$
      & $+0.183435$ & $59.172$            & $18.134$ \\
    5 & $-0.195090$ & $40.245$            & $19.258$
      & $-0.183435$ & $40.828$            & $18.134$ \\
    6 & $-0.555570$ & $22.221$            & $16.326$
      & $-0.525532$ & $23.723$            & $15.685$ \\
    7 & $-0.831470$ & $\phantom{0}8.427$ & $10.909$
      & $-0.796666$ & $10.167$            & $11.119$ \\
    8 & $-0.980785$ & $\phantom{0}0.961$ & $\phantom{0}3.831$
      & $-0.960290$ & $\phantom{0}1.986$ & $\phantom{0}5.061$ \\
    \bottomrule
  \end{tabular}
\end{table}

Applying Gaussian quadrature to the time integral in \cref{eq:A_spectral} gives the following $(\ell+1)$-th Picard iterate for
the acid field, with an analogous expression for the quencher:
\begin{equation}
\begin{aligned}
  \acid^{(\ell+1)}(\mathbf{x},t)
  &= \mathbf{F}_{xy}^{-1}\mathbf{T}_{z}^{-1}
     \bigl[\hat{\mathbf{K}}_A(\mathbf{k},t)\hat{A}_0(\mathbf{k})\bigr](\mathbf{x}) \\
  &\quad - \sum_{i=1}^{N} w_i(t)\,
     \mathbf{F}_{xy}^{-1}\mathbf{T}_{z}^{-1}
     \Bigl[
       \hat{\mathbf{K}}_A(\mathbf{k},t-s_i(t))
       \mathbf{F}_{xy}\mathbf{T}_{z}
        \Bigl[
        {R}^{(\ell)}_A(\mathbf{x},s_i(t))
        \Bigr]
     \Bigr](\mathbf{x}),
\end{aligned}
\label{eq:picard_spectral}
\end{equation}
where the acid reaction source is defined as
\begin{equation}
  {R}^{(\ell)}_A(\mathbf{x},\tau)
  =
  k_3\bigl(A^{(\ell)}(\mathbf{x},\tau)\bigr)^r
  + k_4 A^{(\ell)}(\mathbf{x},\tau)Q^{(\ell)}(\mathbf{x},\tau).
\end{equation}
Within the Picard iteration, the acid field is updated until convergence.

Now the continuous integration over time is turned into the summation over the discrete quadrature nodes. 
To evaluate the concentration fields at the time-dependent quadrature nodes, we represent the complete trajectory of each field using a truncated Chebyshev or Legendre polynomial expansion.
In this way, the concentration fields can be efficiently evaluated at any required time point.
With the normalized time coordinate
\begin{equation}
  \eta(t)=\frac{2t}{t_{PEB}}-1\in[-1,1],
  \label{eq:normalized_time_coordinate}
\end{equation}
the temporal basis functions are
\begin{equation}
  \Phi_j(\eta)
  =
  \begin{cases}
    T_j(\eta)=\cos\!\left(j\arccos\eta\right),
      & \text{Chebyshev basis},\\
    P_j(\eta)=
      \dfrac{1}{2^j j!}
      \dfrac{\mathrm{d}^j}{\mathrm{d}\eta^j}
      \left[(\eta^2-1)^j\right],
      & \text{Legendre basis},
  \end{cases}
  \qquad j=0,\ldots,K-1.
  \label{eq:time_basis_def}
\end{equation}
The complete trajectory is then represented as
\begin{equation}
  u(\mathbf{x},t)
  \approx
  \sum_{j=0}^{K-1}
  c_j^u(\mathbf{x})\,
  \Phi_j\!\left(\eta(t)\right),
  \qquad
  t\in[0,t_{PEB}],
  \qquad
  u\in\{\acid,\quench\},
  \label{eq:global_expansion}
\end{equation}
where $\{c_j^u(\mathbf{x})\}_{j=0}^{K-1}$ are the spatially resolved
temporal coefficient fields and $K$ is the number of the truncated temporal
basis functions.

To determine the $K$ temporal coefficients and thereby the complete solution trajectory at each iteration, the $(\ell+1)$-th Picard update is evaluated at the $K$ temporal projection nodes $\{\tau_i\}_{i=0}^{K-1}$. 
The resulting values are collected as
\begin{equation}
  \mathbf{u}_{\mathrm{Picard}}^{(\ell+1)}(\mathbf{x})
  =
  \bigl(
    u_{\mathrm{Picard}}^{(\ell+1)}(\mathbf{x},\tau_0),
    \ldots,
    u_{\mathrm{Picard}}^{(\ell+1)}(\mathbf{x},\tau_{K-1})
  \bigr)^{\mathsf T}
  \in\mathbb{R}^{K},
  \qquad
  u\in\{\acid,\quench\}.
  \label{eq:picard_rhs}
\end{equation}
These values define the raw Picard coefficient update
\begin{equation}
  \widetilde{\mathbf{c}}^{u,(\ell+1)}(\mathbf{x})
  =
  \left(
    \tilde{c}_{0}^{u,(\ell+1)}(\mathbf{x}),
    \ldots,
    \tilde{c}_{K-1}^{u,(\ell+1)}(\mathbf{x})
  \right)^{\mathsf T}
  \in\mathbb{R}^{K},
  \label{eq:raw_coefficient_vector}
\end{equation}
which is obtained by applying the inverse of the square temporal collocation matrix,
\begin{equation}
  \widetilde{\mathbf{c}}^{u,(\ell+1)}(\mathbf{x})
  =
  \boldsymbol{\Phi}^{-1}
  \mathbf{u}_{\mathrm{Picard}}^{(\ell+1)}(\mathbf{x}),
  \qquad
  u\in\{\acid,\quench\}.
  \label{eq:coeff_solve}
\end{equation}
Here, $\boldsymbol{\Phi}\in\mathbb{R}^{K\times K}$ is the temporal
collocation matrix obtained by evaluating the basis functions at the
projection times,
\begin{equation}
  [\boldsymbol{\Phi}]_{ij}
  =
  \Phi_j\!\left(\eta(\tau_i)\right),
  \qquad
  i,j=0,\ldots,K-1.
  \label{eq:collocation_matrix}
\end{equation}

The collocation solve provides the raw Picard coefficient update.
Anderson acceleration combines this update with recent coefficient iterates to obtain $\mathbf{c}^{u,(\ell+1)}$. 
The updated coefficients are then used to evaluate the concentration fields required for the next Picard iteration. 
The iteration is terminated when the prescribed residual tolerance is reached, at which point the complete trajectories are
determined by the temporal coefficient fields $\mathbf{c}^{A}(\mathbf{x})$ and $\mathbf{c}^{Q}(\mathbf{x})$.
Once the global-in-time Picard iteration for $A$ and $Q$ has converged, all concentration fields can be evaluated at any time within the bake interval using the converged temporal coefficients, without an additional Picard iteration or sequential time stepping.
The coefficient iteration is initialized using the diffusion-only trajectory. 
This trajectory is projected onto the temporal basis after omitting the nonlinear reaction and acid--quencher coupling terms.

In this representation, a small set of temporal coefficient fields replaces the concentration trajectory densely sampled in time at each spatial location, reducing the data volume and computational burden of the coupled PEB system.

\subsection{Mixed-transform Factorized Fourier Neural Operator}
\label{sec:fno_backbone}

The global-in-time coefficient representation used by the numerical method is carried over to the PINO, which predicts temporal coefficient fields rather than concentration fields at individual time locations.
Inspired by the f-FNO architecture~\cite{tran2023factorized}, we adopt a mixed-transform variant of the f-FNO.
This construction replaces a fully coupled three-dimensional spectral kernel with an axial operator followed by factorized lateral operators.
It retains nonlocal interactions along all spatial axes while reducing the parameterization of the spectral kernel.

Let $\mathbf{h}^{(\ell)} \in\mathbb{R}^{d\times N_x\times N_y\times N_z}$ denote the hidden feature field at layer $\ell$.
The axial spectral operator is defined as
\begin{equation}
	\mathcal{S}_{o,z}^{(\ell)}
	\!\left[\mathbf{h}^{(\ell)}\right]
	=
	\mathbf{T}_{z}^{-1}\!\left[
		\sum_{i=1}^{d}
		\mathcal{K}_{oi,z}^{(\ell)}(k_z)
		\mathbf{T}_{z}\!\left[
			h_i^{(\ell)}(x,y,\cdot)
		\right]
	\right](z).
	\label{eq:ffno_z_spectral}
\end{equation}
The axially filtered feature is then
\begin{equation}
	h_{o,z}^{(\ell)}(\mathbf{x})
	=
	\sigma\!\left(
		\mathcal{S}_{o,z}^{(\ell)}
		\!\left[\mathbf{h}^{(\ell)}\right]
		+
		\mathcal{W}_{o,z}^{(\ell)}
		\!\left[\mathbf{h}^{(\ell)}\right]
	\right),
	\label{eq:ffno_z_update}
\end{equation}
where $\sigma$ is the GeLU activation.
The resulting feature $\mathbf{h}_{z}^{(\ell)}$ is subsequently processed by the factorized lateral operator
\begin{equation}
\begin{aligned}
	\mathcal{S}_{o,xy}^{(\ell)}
	\!\left[\mathbf{h}_{z}^{(\ell)}\right]
	={}&
	\mathbf{F}_{x}^{-1}\!\left[
		\sum_{i=1}^{d}
		\mathcal{K}_{oi,x}^{(\ell)}(k_x)
		\mathbf{F}_{x}\!\left[
			h_{i,z}^{(\ell)}(\cdot,y,z)
		\right]
	\right](x)
	\\
	&+
	\mathbf{F}_{y}^{-1}\!\left[
		\sum_{i=1}^{d}
		\mathcal{K}_{oi,y}^{(\ell)}(k_y)
		\mathbf{F}_{y}\!\left[
			h_{i,z}^{(\ell)}(x,\cdot,z)
		\right]
	\right](y).
\end{aligned}
	\label{eq:ffno_xy_spectral}
\end{equation}
The transformed feature and residual update are given by
\begin{align}
	\tilde{h}_{o}^{(\ell)}(\mathbf{x})
	&=
	\sigma\!\left(
		\mathcal{S}_{o,xy}^{(\ell)}
		\!\left[\mathbf{h}_{z}^{(\ell)}\right]
		+
		\mathcal{W}_{o,xy}^{(\ell)}
		\!\left[\mathbf{h}_{z}^{(\ell)}\right]
	\right),
	\\
	h_o^{(\ell+1)}(\mathbf{x})
	&=
	h_o^{(\ell)}(\mathbf{x})
	+
	\tilde{h}_{o}^{(\ell)}(\mathbf{x}),
	\qquad o=1,\ldots,d.
	\label{eq:ffno_2d_update}
\end{align}
Here, $d$ is the number of hidden channels, $h_o^{(\ell)}$ denotes the
$o$-th hidden channel at layer $\ell$, and
$\mathcal{W}_{\alpha}^{(\ell)}[\mathbf{h}]
=W_{\alpha}^{(\ell)}\mathbf{h}+b_{\alpha}^{(\ell)}$,
$\alpha\in\{z,xy\}$, denotes an affine $1\times1\times1$ convolution that
mixes the hidden channels. 

For training, the mixed-transform f-FNO predicts the temporal coefficient fields
$\mathbf{c}_{\theta}^{u}$ of the nonlinear correction over the full bake
interval $[0,t_{\mathrm{PEB}}]$.
The total acid and quencher trajectories are reconstructed by combining the
analytical diffusion-only base solution with the predicted correction,
\begin{equation}
\begin{aligned}
	u_{\theta}(\mathbf{x},t)
	&= u_{\mathrm{base}}(\mathbf{x},t)
		+ u_{\mathrm{corr},\theta}(\mathbf{x},t), \\
	u_{\mathrm{corr},\theta}(\mathbf{x},t)
	&= \sum_{j=0}^{n_c-1}
		c_{\theta,j}^u(\mathbf{x})\,
		\Phi_j\!\left(\eta(t)\right),
	\qquad
	\eta(t)=\frac{2t}{t_{\mathrm{PEB}}}-1,
	\qquad
	u\in\{\acid,\quench\}.
\end{aligned}
	\label{eq:pino_coefficient_reconstruction}
\end{equation}
Here, $u_{\mathrm{base}}$ is the analytical diffusion-only solution,
$c_{\theta,j}^u(\mathbf{x})$ is the $j$-th coefficient field predicted by
the mixed-transform f-FNO, and $n_c$ is the temporal truncation count for the nonlinear
correction, analogous to $K$ in \cref{eq:global_expansion}.

For fixed diffusion coefficients and the reaction coefficients
$k_1,\ldots,k_5$ in \cref{eq:P,eq:D,eq:A,eq:Q}, the mixed-transform f-FNO defines the
operator
\begin{equation}
	\mathcal{G}_{\theta}:
	\left(\acid_0,\quench_0,x,y,z\right)
	\longmapsto
	\left(
		\mathbf{c}_{\theta}^{A},
		\mathbf{c}_{\theta}^{Q}
	\right),
	\qquad
	\mathbf{c}_{\theta}^{u}
	= \left\{c_{\theta,j}^u\right\}_{j=0}^{n_c-1},
	\quad
	u\in\{\acid,\quench\}.
	\label{eq:pino_operator_map}
\end{equation}
The input fields and three coordinate channels are concatenated and lifted
by a $1\times1\times1$ convolution to $d$ hidden channels.
With the analytical base solution enabled, the physical concentration
fields are reconstructed according to
\cref{eq:pino_coefficient_reconstruction}, and the PEB
differential-residual and Duhamel-consistency losses are evaluated on the
total reconstructed fields $u_{\theta}$.
As in the spectral-in-time Picard method, the protected and deprotected
fields are evaluated during inference from the predicted acid
trajectory.

\subsection{Loss functions for mixed-transform f-FNO PINO}
\label{sec:pde_loss}

At every training iteration, $\mathcal{G}_{\theta}$ predicts the temporal coefficient fields $\mathbf{c}_{\theta}^{A}$ and $\mathbf{c}_{\theta}^{Q}$ defined in \cref{eq:pino_operator_map}. 
The reconstructed fields $u_{\theta}$ in \cref{eq:pino_coefficient_reconstruction} are then evaluated at temporal collocation nodes $\{t_n\}_{n=1}^{N_t}\subset[0,t_{PEB}]$.

We normalize the differential residual loss by the squared initial-field
magnitude to balance the acid and quencher contributions,
\begin{equation}
	\mathcal{L}_{\mathrm{PDE}}^u
	= \frac{1}{N_t\,\overline{u_0^2}}
		\sum_{n=1}^{N_t}
		\left\|\mathcal{R}_u[u_{\theta}](\cdot,t_n)\right\|_2^2,
	\qquad u\in\{\acid,\quench\},
	\label{eq:pino_pde_loss}
\end{equation}
where $\overline{u_0^2}$ denotes the spatial mean of $u_0^2$ and $\|\cdot\|_2^2$ denotes the spatially averaged squared norm. 
The quantity $\mathcal{R}_u$ denotes the differential residual of the corresponding governing equation as defined by,
\begin{align}
  \mathcal{R}_A
  &=
  \partial_t A_{\theta}
  + k_3 A_{\theta}^{r}
  + k_4 A_{\theta}Q_{\theta}
  - D_A\nabla^2 A_{\theta},
  \\
  \mathcal{R}_Q
  &=
  \partial_t Q_{\theta}
  + k_5 Q_{\theta}^{s}
  + k_4 A_{\theta}Q_{\theta}
  - D_Q\nabla^2 Q_{\theta}.
\end{align}
The temporal derivative of the reconstructed field is evaluated analytically as
\begin{equation}
	\partial_t u_{\theta}(\mathbf{x},t)
	= \partial_t u_{\mathrm{base}}(\mathbf{x},t)
		+ \sum_{j=0}^{n_c-1}
		c_{\theta,j}^u(\mathbf{x})\,
		\Phi_j'\!\left(\eta(t)\right)\frac{2}{t_{PEB}},
	\qquad u\in\{\acid,\quench\},
	\label{eq:pino_temporal_derivative}
\end{equation}
Spatial Laplacians are evaluated using the same pseudo-spectral formulation. 

To construct the Duhamel integral-consistency losses
$\mathcal{L}_{\mathrm{int}}^A$ and $\mathcal{L}_{\mathrm{int}}^Q$, we form
the set $\mathcal{P}$ of all chronological pairs of the temporal
collocation nodes,
\begin{equation}
	\mathcal{P}
	= \left\{(t_i,t_j)\,\middle|\,1\leq i<j\leq N_t\right\},
	\qquad
	N_{\mathrm{pair}}
	= \lvert\mathcal{P}\rvert
	= \frac{N_t(N_t-1)}{2}.
	\label{eq:pino_duhamel_time_pairs}
\end{equation}
For each pair $(t_i,t_j)\in\mathcal{P}$,
$\mathcal{I}_u[u_{\theta}](\cdot;t_i,t_j)$ denotes the residual of the
Duhamel integral relation introduced in
Section~\ref{sec:Duhamel_spectral} over $[t_i,t_j]$. The integral is
evaluated using the adopted temporal quadrature rule, and the corresponding
integral-consistency loss is defined as
\begin{equation}
	\mathcal{L}_{\mathrm{int}}^u
	= \frac{1}{N_{\mathrm{pair}}\,\overline{u_0^2}}
		\sum_{(t_i,t_j)\in\mathcal{P}}
		\left\|
			\mathcal{I}_u[u_{\theta}](\cdot;t_i,t_j)
		\right\|_2^2,
	\qquad
	u\in\{\acid,\quench\}.
	\label{eq:pino_integral_loss}
\end{equation}
In addition to \cref{eq:pino_integral_loss}, the Duhamel residual over the
full bake interval $(0,t_{\mathrm{PEB}})$ is evaluated separately at every
training iteration to form the endpoint integral-consistency losses.

The initial-condition term enforces the reconstructed fields at $t=0$ to match $(\acid_0,\quench_0)$,
\begin{equation}
	\mathcal{L}_{\mathrm{IC}}
	= \sum_{u\in\{\acid,\quench\}}
		\left\|u_{\theta}(\cdot,0)-u_0\right\|_2^2.
	\label{eq:pino_initial_loss}
\end{equation}

The differential and integral residuals at the final bake time, together with the corresponding non-negativity penalties, are collected into the endpoint loss,
\begin{equation}
	\mathcal{L}_{\mathrm{end}}
	= \sum_{u\in\{\acid,\quench\}}
		\left(
			\lambda_{\mathrm{PDE},t_{\mathrm{PEB}}}^u
			\mathcal{L}_{\mathrm{PDE},t_{\mathrm{PEB}}}^u
			+
			\lambda_{\mathrm{int},t_{\mathrm{PEB}}}^u
			\mathcal{L}_{\mathrm{int},t_{\mathrm{PEB}}}^u
			+
			\lambda_{\mathrm{neg},t_{\mathrm{PEB}}}^u
			\mathcal{L}_{\mathrm{neg},t_{\mathrm{PEB}}}^u
		\right).
	\label{eq:pino_endpoint_loss}
\end{equation}
Here, $\mathcal{L}_{\mathrm{PDE},t_{\mathrm{PEB}}}^u$ denotes the differential residual loss evaluated at $t=t_{\mathrm{PEB}}$, $\mathcal{L}_{\mathrm{int},t_{\mathrm{PEB}}}^u$ denotes the Duhamel integral residual loss evaluated over $[0,t_{\mathrm{PEB}}]$, and 
\begin{equation}
	\mathcal{L}_{\mathrm{neg},t_{\mathrm{PEB}}}^u
	=
	\left\|
		\max\left(
			0,-u_{\theta}(\cdot,t_{\mathrm{PEB}})
		\right)
	\right\|_2^2
	\label{eq:pino_endpoint_negativity}
\end{equation}
penalizes negative concentrations at the final bake time.

In addition, we penalize negative concentrations and total-mass increases over time. 
These constraints are collected as
\begin{equation}
	\mathcal{L}_{\mathrm{con}}
	= \sum_{m\in\mathcal{C}}\lambda_m\mathcal{L}_m,
	\label{eq:pino_constraint_loss}
\end{equation}
where $\mathcal{C}$ denotes the set of auxiliary physical constraints. 

The physics-informed objective is a weighted sum,
\begin{equation}
\begin{aligned}
	\mathcal{L}
	={}& \lambda_{\mathrm{PDE},A}\mathcal{L}_{\mathrm{PDE}}^A
		+ \lambda_{\mathrm{PDE},Q}\mathcal{L}_{\mathrm{PDE}}^Q
		+ \lambda_{\mathrm{int},A}\mathcal{L}_{\mathrm{int}}^A
		+ \lambda_{\mathrm{int},Q}\mathcal{L}_{\mathrm{int}}^Q \\
	 &+ \lambda_{\mathrm{IC}}\mathcal{L}_{\mathrm{IC}}
		+ \mathcal{L}_{\mathrm{end}}
		+ \mathcal{L}_{\mathrm{con}}.
	\label{eq:pino_total_loss}
\end{aligned}
\end{equation}

\section{Results and Discussion}
\label{sec:results}
This section evaluates the proposed spectral Duhamel--Picard method for global-in-time solution of the three-dimensional PEB problem. 
We first examine convergence with polynomial order to determine the temporal expansion order required to represent the diffusion- and reaction-driven evolution accurately. 
We then assess the convergence and reliability of the method in three-dimensional PEB simulations. 
The PINO is trained to minimize the physics-informed loss defined in \cref{eq:pino_total_loss}, with the same global-in-time temporal expansion and gaussian quadrature used in the spectral Duhamel--Picard formulation shown in \cref{sec:chebyshev_picard}.
The performance of the PINO is subsequently evaluated through its training and inference results. 
The concentration fields are further used to evaluate the topological characteristics of the resulting patterns after development process.
Finally, we compare the proposed approach with conventional time-integration methods to quantify its computational efficiency.
All numerical computations, including PINO training and inference, were performed on a single NVIDIA A100 GPU with 80~GB of memory.

\subsection{Temporal Polynomial Convergence}
\label{sec:results_polynomial_convergence}
\begin{figure}[H]
\centering
\includegraphics[width=0.3\textwidth]{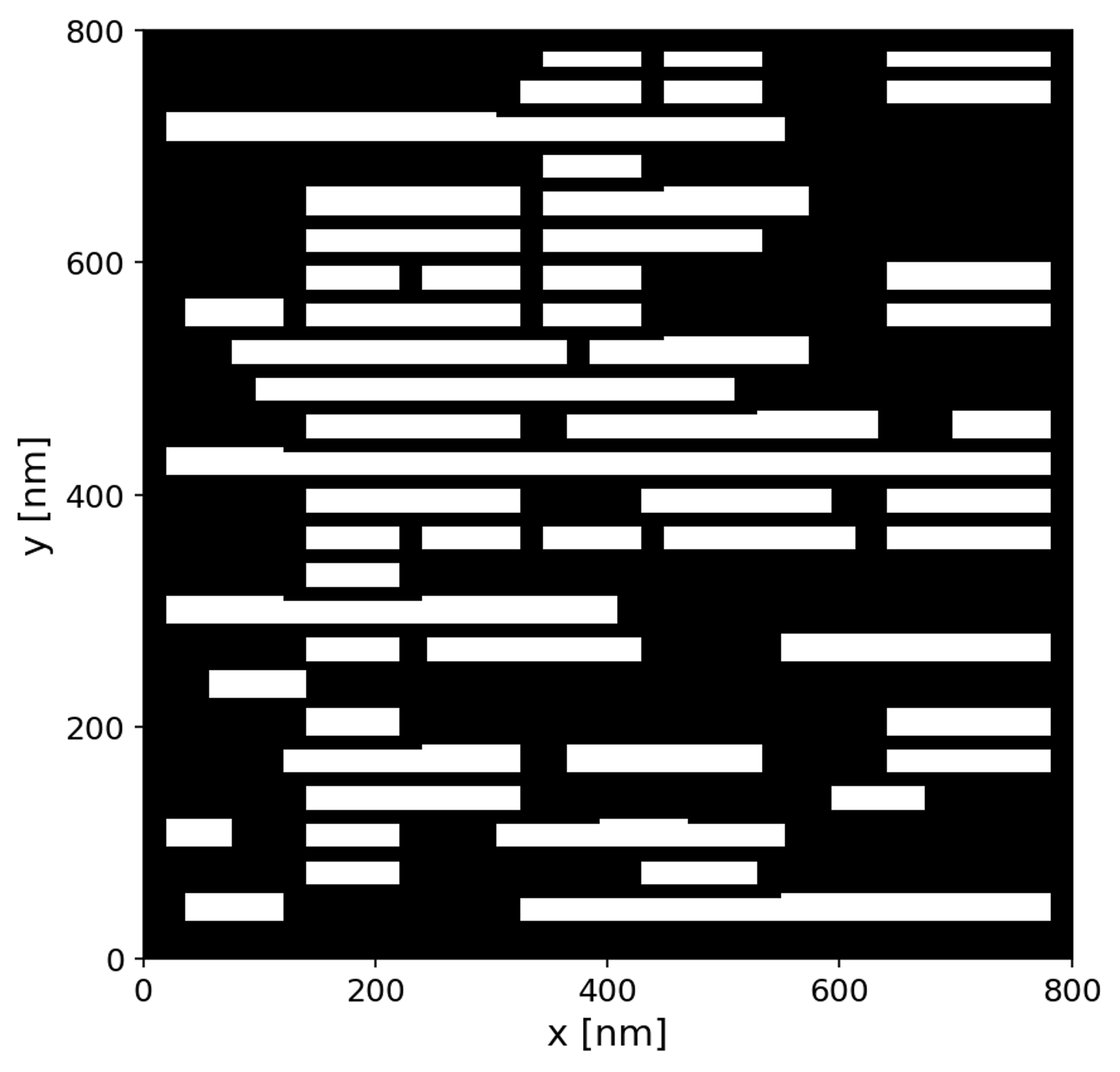}
\caption{Random binary mask containing horizontally oriented features, used to construct the initial acid distribution in the $800\times 800\times 200$ PEB benchmark domain.}
\label{fig:results_mask_layout}
\end{figure}

We first solve the three-dimensional PEB problem using a conventional fourth-order Runge--Kutta (RK4) time-marching scheme combined with a fourth-order finite difference (FD4) to obtain a reference trajectory.
For this benchmark, the optical image-formation steps are omitted, and the initial acid distribution is constructed from a random binary mask containing horizontally oriented features on an $800\times 800\times 200$ computational domain, as shown in \cref{fig:results_mask_layout}.

We begin by defining the lateral distribution of the initial acid concentration from the binary mask, assigning $A_{0,\mathrm{in}}=0.5$ to the transmitting regions and $A_{0,\mathrm{out}}=0$ to the opaque regions at the resist top surface.
An effective depth dependence is then prescribed as $A_0(x,y,z)=A_0(x,y,z_{\mathrm{top}}) \exp[-\alpha_A(z_{\mathrm{top}}-z)]$, where $\alpha_A$ controls the attenuation of the initial acid concentration with depth. 
The initial acid field is subsequently Gaussian-blurred as a practical surrogate for lateral optical low-pass filtering and for suppressing near-Nyquist components on the pixelated computational grid. 
The initial quencher concentration is uniformly set to $Q_0(x,y,z)=0.05$.
The fixed initial conditions and parameter ranges considered in this evaluation are listed in \cref{tab:results_parameters}.
\begin{table}[H]
\centering
\caption{Fixed initial conditions and parameter ranges used in the temporal fitting benchmark.}
\label{tab:results_parameters}
\begin{tabular}{lcc}
\toprule
Parameter & Description & Value \\
\midrule
$A_0$ & Inside / outside mask & $0.5$ / $0.0$ \\
$Q_0$ & Uniform initial quencher concentration & $0.05$ \\
$P_0$ & Uniform initial protected-group concentration & $1.0$ \\
$D_0$ & Uniform initial deprotected-group concentration & $0.0$ \\
$k_1,\,k_4$ & Reaction rate constants & $0.01\text{--}1$ \\
$k_2,\,k_3,\,k_5$ & Reaction rate constants & $0.0001\text{--}0.01$ \\
$p,\,q,\,r,\,s$ & Reaction exponents & $0.5\text{--}2.0$ \\
$D_{\mathrm{acid}}, D_{\mathrm{q}}$ & Acid and quencher diffusion coefficients & $1\text{--}20~\mathrm{nm}^2/\mathrm{s}$ \\
$D_d$ & Deprotected-group diffusion coefficient & $0.05\text{--}1~\mathrm{nm}^2/\mathrm{s}$ \\
$t$ & Temporal domain & $0\le t\le 100~\mathrm{s}$ \\
\bottomrule
\end{tabular}
\end{table}

\begin{figure}[H]
\centering
\includegraphics[width=1.0\textwidth]{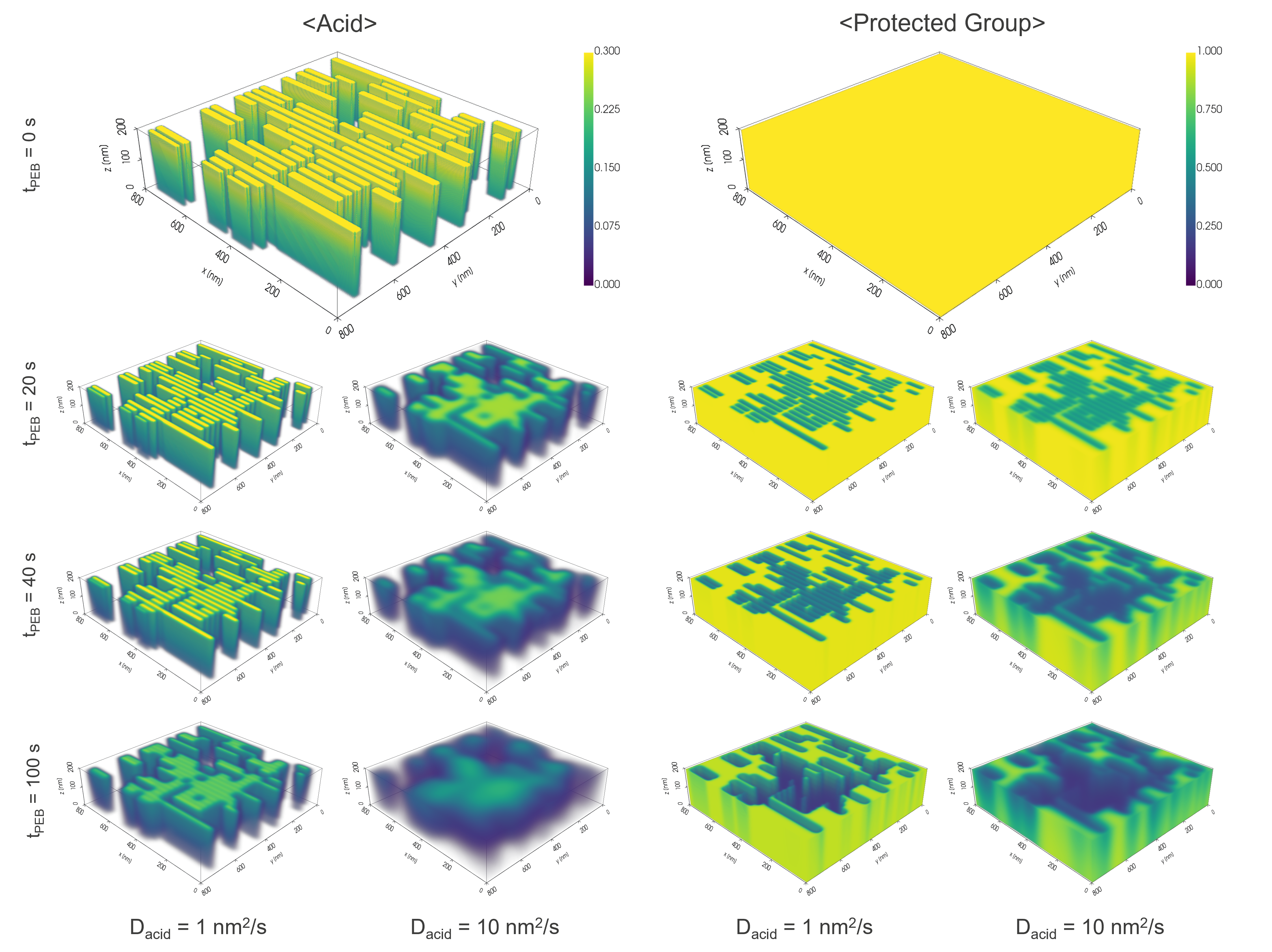}
\caption{Reference time-marching trajectories for acid ($A$) and protected group ($P$) over $t=0$ to $100$ s, with representative acid diffusion coefficients $D_{\mathrm{acid}}=1$ and $10~\mathrm{nm}^2/\mathrm{s}$.}
\label{fig:results_peb_simulation}
\end{figure}

Using representative parameter values selected from these ranges, the coupled reaction--diffusion system is solved by time marching over the PEB interval $t\in[0,100]~\mathrm{s}$.
Concentration fields at selected times are presented in \cref{fig:results_peb_simulation}, highlighting the acid ($A$) and protected-group ($P$) fields for two representative matched acid and quencher diffusion settings, $D_{\mathrm{acid}}=D_{\mathrm{q}}\in\{1,10\}~\mathrm{nm}^2/\mathrm{s}$.

\begin{figure}[H]
\centering
\includegraphics[width=1.0\textwidth]{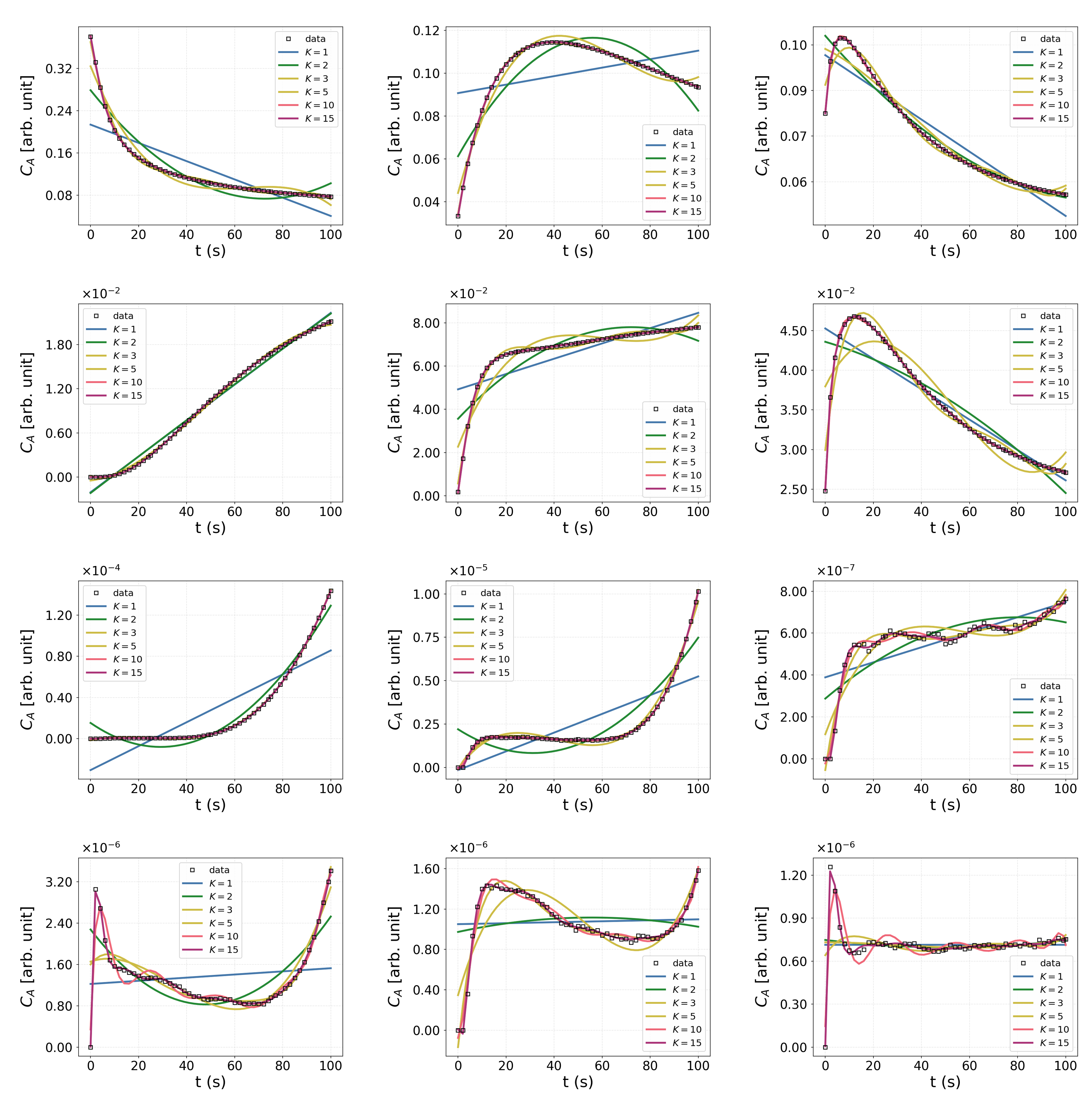}
\caption{Representative temporal fitting of acid trajectories at sampled voxel locations selected to cover diverse local profile shapes.}
\label{fig:results_fitting}
\end{figure}

From the time-marching reference in \cref{fig:results_peb_simulation}, concentration fields are stored at discrete times $t_i$ ($i=1,\dots,N_t$).
Accordingly, each voxel contains a temporal sequence of length $N_t$.
For each voxel, the acid-concentration sequence is fitted by a temporal polynomial, and the fitting agreement with the time-marching reference is evaluated as a function of polynomial order $K$.
Representative fitting examples for acid trajectories at sampled voxel locations are shown in \cref{fig:results_fitting}.
The selected samples represent diverse local trajectory shapes and magnitudes across different spatial contexts.
With increasing polynomial order $K$, the fitted concentration histories approach the time-marching references, and beyond $K=10$ the difference from the reference acid trajectories is visually indistinguishable.
As $K$ increases, the coefficient of determination ($R^2$) approaches unity and the RMSE relative to the time-marching reference decreases approximately exponentially.
These results demonstrate that the polynomial expansion closely captures the temporal concentration trajectories with higher-order polynomials providing increasingly accurate representations.
The coefficient-magnitude analysis further reveals a substantial decay beyond orders 4--5, suggesting that the higher-order modes primarily provide small refinements to the reconstructed trajectories.

For a polynomial order $K$, the coefficient tensor has size $N_x\times N_y\times N_z\times K$, replacing the full snapshot storage $N_x\times N_y\times N_z\times N_t$ while preserving trajectory reconstruction accuracy.
The $N_t$ varies depending on the temporal resolution of the time-marching reference, but the polynomial expansion with order $K$ provides a more compact representation when $K \ll N_t$.

\begin{figure}[H]
    \centering
    \begin{tikzpicture}
        \node[anchor=south west, inner sep=0] (image) at (0,0)
            {\includegraphics[width=1.0\textwidth]{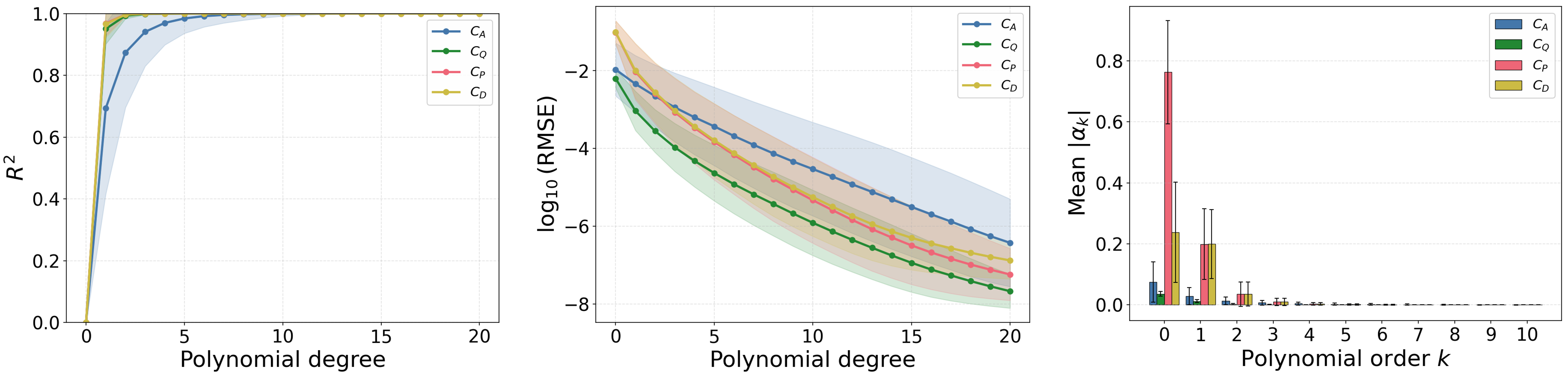}};
        \begin{scope}[
            shift={(image.south west)},
            x={(image.south east)},
            y={(image.north west)}
        ]
            \node[anchor=north west, font=\bfseries, inner sep=1.5pt]
                at (0.00,1.1) {(a)};
            \node[anchor=north west, font=\bfseries, inner sep=1.5pt]
                at (0.34,1.1) {(b)};
            \node[anchor=north west, font=\bfseries, inner sep=1.5pt]
                at (0.67,1.1) {(c)};
        \end{scope}
    \end{tikzpicture}
    \caption{
        Global fitting statistics over all voxel locations for the $A$, $Q$,
        $P$, and $D$ fields:
        (a) coefficient of determination $R^2$,
        (b) trajectory mismatch relative to the time-marching reference, and
        (c) normalized coefficient contributions for polynomial expansions
        with different orders $K$.
    }
    \label{fig:results_fitting2}
\end{figure}

\subsection{Global-in-time spectral Duhamel--Picard PEB}
\label{sec:results_global_in_time}

\begin{figure}[H]
\centering
\includegraphics[width=1.0\textwidth]{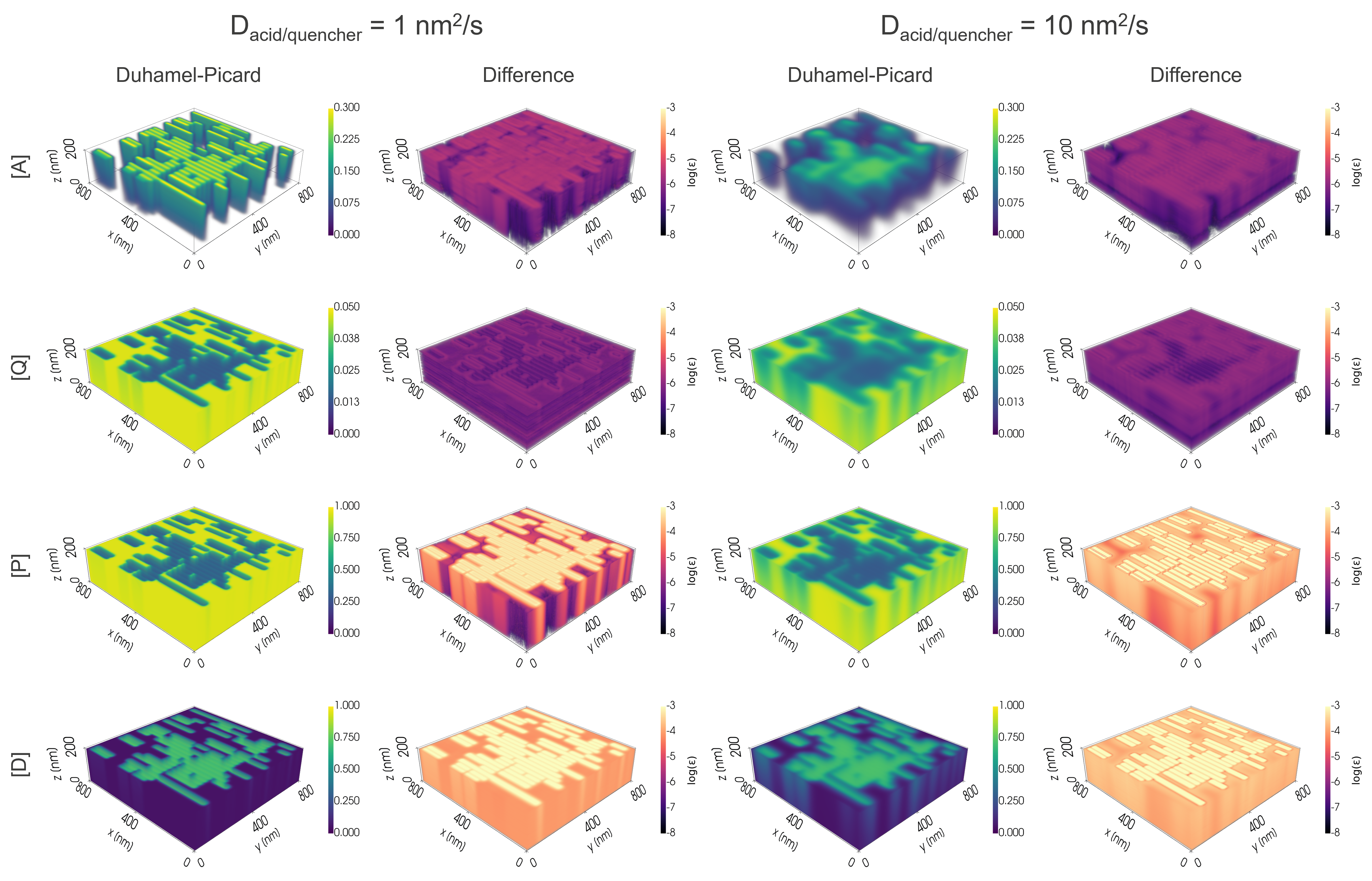}
\caption{Global-in-time spectral Duhamel--Picard solutions for the acid ($A$), quencher ($Q$), protected-group ($P$), and deprotected-group ($D$) fields at $t_{PEB}=50~\mathrm{s}$ for representative diffusion settings with $D_{\mathrm{acid}}=D_{\mathrm{q}}\in\{1,10\}~\mathrm{nm}^2/\mathrm{s}$. The corresponding absolute differences relative to the FD4--RK4 time-marching reference obtained using $\Delta t=0.01~\mathrm{s}$ are shown alongside the computed fields.}
\label{fig:results_global_in_time}
\end{figure}

We next evaluate the proposed global-in-time spectral Duhamel--Picard method for the 3-D PEB reaction--diffusion system.
Unless otherwise specified, a polynomial order of $K=10$ and a quadrature order of $N=8$ are used throughout this subsection.
\Cref{fig:results_global_in_time} presents the solutions for the $A$, $Q$, $P$, and $D$ fields evaluated at $t_{PEB}=50~\mathrm{s}$ for representative diffusion settings with $D_{\mathrm{acid}}=D_{\mathrm{q}}\in\{1,10\}~\mathrm{nm}^2/\mathrm{s}$.

\noindent With these settings, the global-in-time Picard iteration for $A$ and $Q$ reaches a self-consistency level of approximately $10^{-7}$ within 10 iterations.
The $P$ and $D$ fields are subsequently updated based on the converged $A$ and $Q$ fields.
Once the global-in-time Picard iteration for $A$ and $Q$ has converged, the all concentration fields can be easily evaluated at any arbitrary time within the computed time using the optimized temporal coefficients without performing an additional Picard iteration or sequential time stepping.

The figure also shows the corresponding absolute differences from the FD4--RK4 time-marching reference computed with $\Delta t=0.01~\mathrm{s}$.
The global-in-time Picard iteration solutions exhibit good agreement with the time-marching reference.
The mean absolute errors (MAEs) between the global-in-time Picard iteration and FD4--RK4 solutions are on the order of $10^{-6}$ for $A$ and $Q$, and on the order of $10^{-4}$ for $P$ and $D$ at the test PEB time $t_{PEB}=50~\mathrm{s}$.
The MAEs for $P$ and $D$ are approximately two orders of magnitude larger than those for $A$ and $Q$ because numerical differences in the acid and protected-group trajectories accumulate through their time-integrated reaction histories.

\begin{figure}[H]
    \centering
    \begin{tikzpicture}
        \node[anchor=south west, inner sep=0] (image) at (0,0)
            {\includegraphics[width=1.0\textwidth]{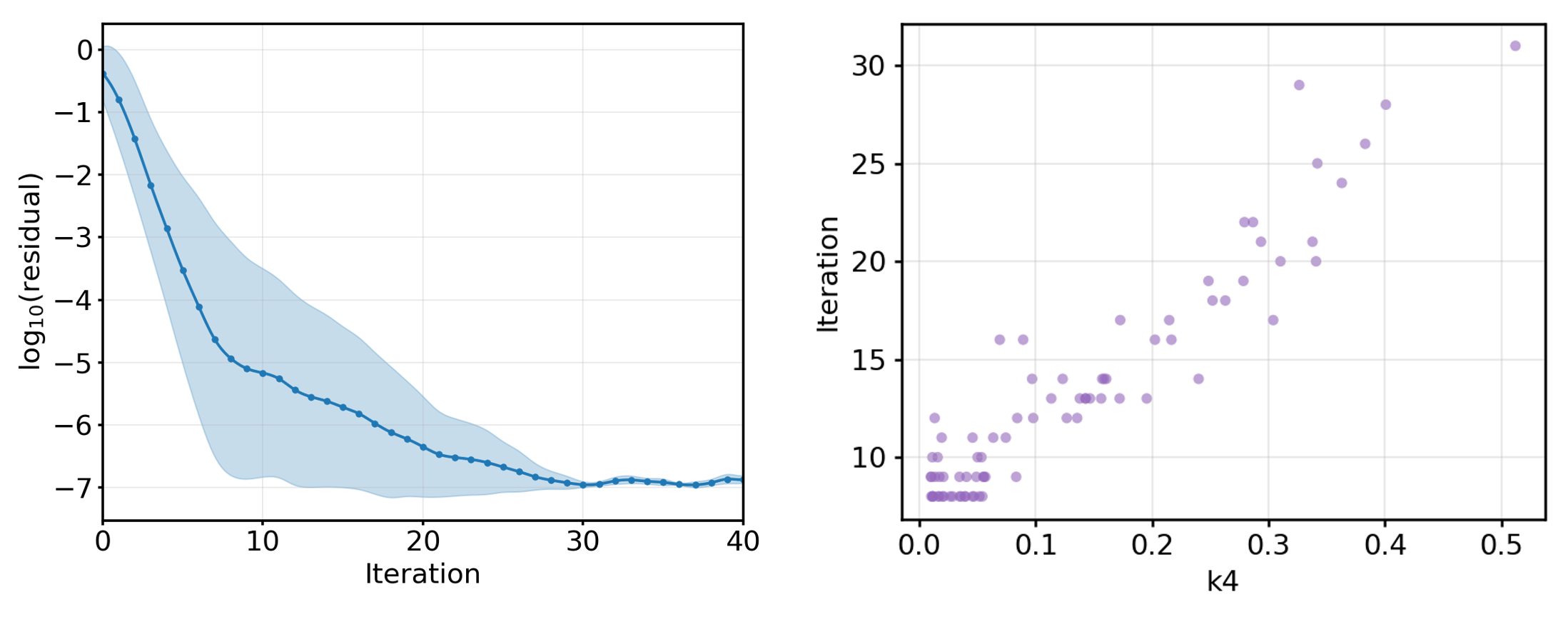}};
        \begin{scope}[
            shift={(image.south west)},
            x={(image.south east)},
            y={(image.north west)}
        ]
            \node[anchor=north west, font=\bfseries, inner sep=1.5pt]
                at (0.05,1.05) {(a)};
            \node[anchor=north west, font=\bfseries, inner sep=1.5pt]
                at (0.55,1.05) {(b)};
        \end{scope}
    \end{tikzpicture}
    \caption{
        Convergence behavior of the global-in-time spectral Duhamel--Picard
        solver for 100 parameter combinations generated by Latin hypercube
        sampling over the ranges listed in
        \cref{tab:results_parameters}.
        (a) Convergence histories of the sampled cases.
        (b) Relationship between $k_4$ and the required number of Picard
        iterations.
    }
    \label{fig:iteration_trends}
\end{figure}

To assess convergence over the parameter ranges listed in \cref{tab:results_parameters}, 100 distinct parameter combinations were generated using Latin hypercube sampling, in order to investigate the parameter space effectively.
The global-in-time Picard iteration was applied independently to each combination using the same polynomial and quadrature orders.
In this evaluation, the domain size of $1200 \times 1200 \times 100$ voxels was used with a grid size of $1~\mathrm{nm}$.
\Cref{fig:iteration_trends} summarizes the resulting convergence behavior.
The mean iteration to reach the residual error below $10^{-5}$ is less than 20. 
Among the sampled input parameters, $k_4$ exhibited the clearest positive linear association with the required number of Picard iterations, as shown in the right panel of \cref{fig:iteration_trends}. 
Within the Duhamel--Picard update, the exact change in the acid--quencher coupling term between two iterations is $k_4(Q\,\delta A+A\,\delta Q+\delta A\,\delta Q)$, where $\delta A$ and $\delta Q$ denote the iteration-to-iteration changes in the two fields. 
A larger $k_4$ therefore produces a larger change in the coupled reaction term for the same $\delta A$ and $\delta Q$ resulting in an increase in the number of iterations required for convergence.
No pronounced trend was observed for the other parameters over the investigated ranges.

The wall-clock performance of the proposed method is compared with that of FD4--RK4 in \cref{tab:results_runtime_comparison} with $D_a=D_q=10~\mathrm{nm}^2/\mathrm{s}$, $k_1=k_4=0.1$, $k_2=k_3=k_5=0.001$, and initial concentrations set to $A_0 = 0.5$, $Q_0 = 0.05$. For this benchmark, the domain size of $1200 \times 1200 \times 100$ voxels was used again with a grid size of $1~\mathrm{nm}$.
The global-in-time calculation takes about $80~\mathrm{s}$ to reach a convergence tolerance of $10^{-5}$.
For FD4--RK4, $\Delta t=0.0125~\mathrm{s}$ was adopted to maintain the evaluated discrepancy below $10^{-5}$ relative to the $\Delta t=0.01~\mathrm{s}$ reference.
With this time step, FD4--RK4 requires approximately $2{,}700~\mathrm{s}$ (or $45~\mathrm{min}$) on the GPU and $221{,}600~\mathrm{s}$ (or $61.6~\mathrm{h}$) on the CPU, as tabulated in \cref{tab:results_runtime_comparison}.
Accordingly, the global-in-time Picard iteration was approximately $33.8$ times faster than the GPU implementation of FD4--RK4 and $2.8\times10^3$ times faster than the CPU implementation, respectively.
Neither the $P$ nor the $D$ field requires Picard iterations. Their computation at an evaluation time $t$ takes approximately $120~\mathrm{ms}$ and $1.2~\mathrm{s}$, respectively. The higher cost for $D$ results from the spectral propagation required within its temporal quadrature, whereas $P$ involves only pointwise temporal integration.

\begin{table}[H]
\centering
\small
\caption{Wall-clock runtime comparison between the global-in-time method and conventional FD4--RK4 method run on CPU and GPU.}

\label{tab:results_runtime_comparison}
\begin{tabular}{llrr}
\toprule
Method & Device & Runtime (s) & $T/T_{\mathrm{Picard}}$ \\
\midrule
FD4--RK4 & CPU & $221{,}600$ & $2.8\times10^3$ \\
FD4--RK4 & GPU & $2{,}700$ & $33.8$ \\
Global-in-time Spectral Duhamel-Picard & GPU & $80$ & $1$ \\
\bottomrule
\end{tabular}
\end{table}

It should also be emphasized that the accuracy and convergence criteria for the Picard iteration differ from those of the conventional FD4--RK4 method.
The FD4--RK4 time step is determined based on the discrepancy from the fine-step reference, whereas the global-in-time Picard iteration is terminated based on iterative self-convergence. 
Thus, the reported runtime ratios do not correspond to an identical error level and may vary with the physical parameters, as slower diffusion or reaction kinetics can permit larger time steps for FD4--RK4, for instance. 
Nevertheless, the results demonstrate the computational advantage of the global-in-time method in cases where FD4--RK4 requires small time steps to maintain accuracy.

\subsection{PINO for Global-in-time PEB}
\label{sec:results_PINO}
In this subsection, we investigate the performance of the PINO model trained with the global-in-time spectral Duhamel PEB formulation.
The neural operator is designed to map the initial $A$ and $Q$ concentration fields to spatially resolved temporal coefficients over the fixed bake interval.
The reconstructed trajectories are then evaluated at randomly sampled times within $[0,100]~\mathrm{s}$ during training and validation.
The temporal coefficients are inferred from the initial concentration fields, and the solution fields can subsequently be reconstructed at any time within the fixed bake interval.

For the training dataset, we use a collection of about ${10,000}$ random binary masks containing horizontally oriented features on a $300\times300\times100$ computational domain with a grid of 4 nm for x- and y-directions and 1 nm for the z-direction.
Evaluation times are randomly sampled from the interval $[0, 100]~\mathrm{s}$, ensuring that the neural operator is exposed to a diverse set of temporal scenarios during training.
The PEB process parameters are fixed at $D_a=D_q=2~\mathrm{nm}^2/\mathrm{s}$, $k_1=k_4=0.05$, $k_2=k_3=k_5=0.001$, and $p$=$q$=$r$=$s$=1. 
Upon convergence, the PDE residual losses defined in Eq.~\eqref{eq:pino_pde_loss} reached approximately \(10^{-5}\) and \(10^{-4}\) for \(A\) and \(Q\), respectively.
For the validation test, we use a separate set of random binary masks that are not included in the training dataset.
The model is trained without any precomputed solutions, relying solely on the aforementioned governing equations and physical constraints to learn the global-in-time mapping.

\begin{figure}[H]
\centering
\includegraphics[width=1.0\textwidth]{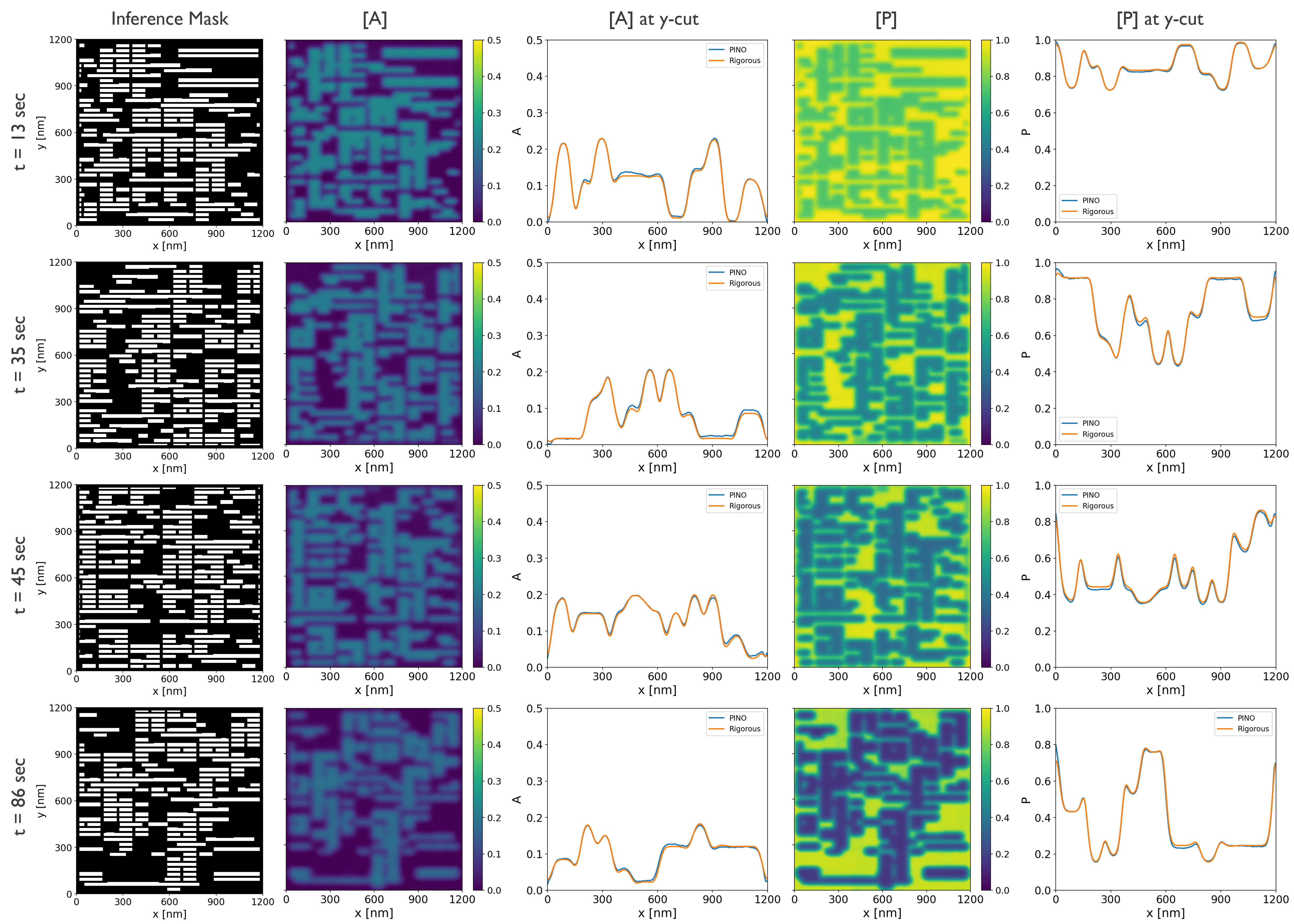}
\caption{Performance of the Physics-Informed Neural Operator (PINO) trained with the global-in-time PEB formulation evaluated for arbitrary validation masks at random bake times, which were not explicitly included in the training dataset. The figure shows the predicted fields and their differences relative to the reference PEB solution.}
\label{fig:results_pino}
\end{figure}

\begin{figure}[H]
    \centering
    \begin{tikzpicture}
        \node[anchor=south west, inner sep=0] (image) at (0,0)
            {\includegraphics[width=0.8\textwidth]{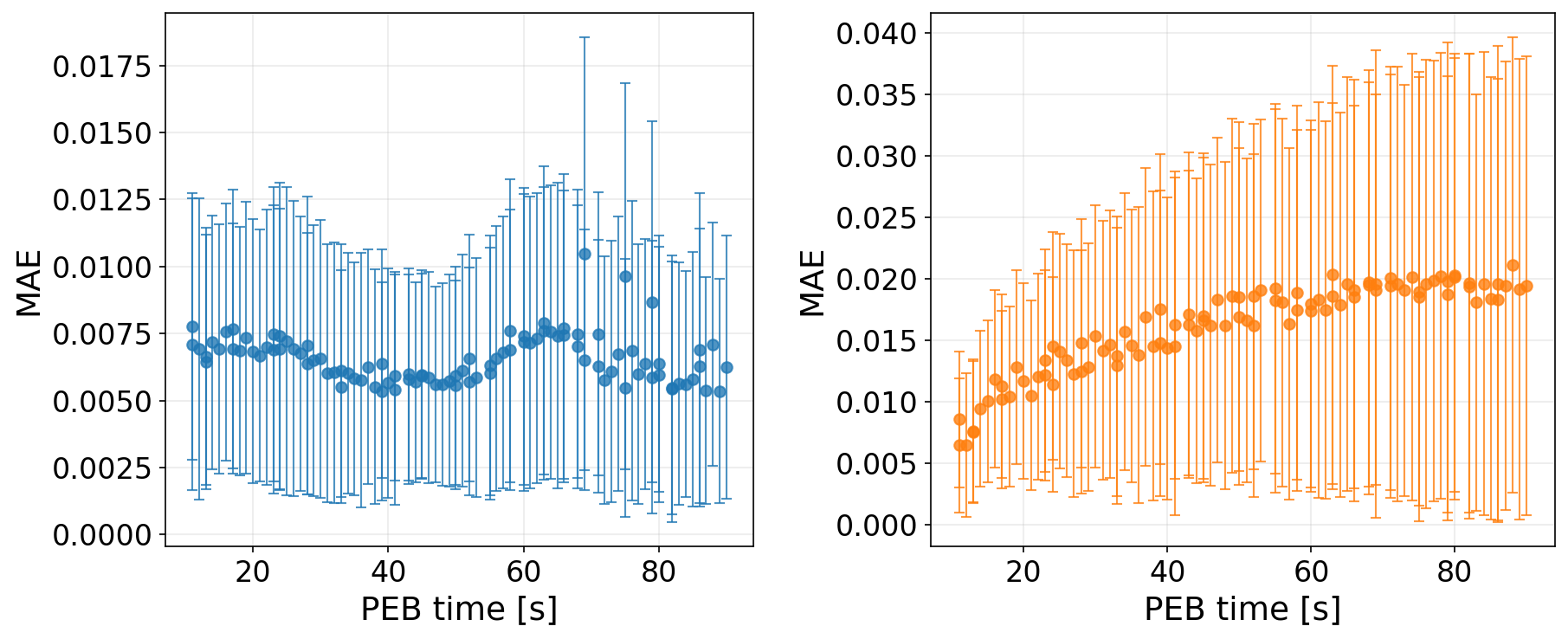}};
        \begin{scope}[
            shift={(image.south west)},
            x={(image.south east)},
            y={(image.north west)}
        ]
            \node[anchor=north west, font=\bfseries, inner sep=1.5pt]
                at (0.05,1.07) {(a)};
            \node[anchor=north west, font=\bfseries, inner sep=1.5pt]
                at (0.52,1.07) {(b)};
        \end{scope}
    \end{tikzpicture}
    \caption{Mean absolute errors for (a) acid and (b) protected groups as a function of bake time.}
    \label{fig:results_mae}
\end{figure}

In \cref{fig:results_pino}, we present the $A$ concentration fields reconstructed from the temporal coefficients predicted by the PINO for various mask patterns and evaluation times.
The corresponding $P$ concentration fields are also shown on the right side, which are computed by \cref{eq:P_linear} using the $A$. 
From the cross-sectional concentration profiles, it is clear that the PINO predictions closely match the reference Picard solutions both for the acid and protected groups.
More detailed comparisons and quantitative evaluations are provided in \cref{fig:results_mae}, where the mean absolute error (MAE) for both the acid and protected groups is plotted as a function of bake time, that was randomly sampled during the validation with random masks.

From this graph, it is observed that the MAE for $A$ remains relatively low across the entire range of bake times, indicating that the PINO inference is reasonably accurate and reliable throughout the temporal domain.
The MAE for $P$, however, tends to increase slightly with bake time.
The increasing discrepancy in $P$ can be attributed to the accumulation of differences in the acid concentration history between the PINO and reference solutions. 
Despite this accumulated error, the overall agreement throughout the bake interval demonstrates the neural operator's ability to capture the global-in-time PEB dynamics represented by the spectral Duhamel formulation.
For this test domain size, the PINO inference time was about 80{\%} of the time required for the reference Picard solution.
Due to memory limitations, the PINO could not be trained directly on the $1200\times1200\times100$ domains used for the reference Picard solutions with the computational resources and implementation employed in this study.

\subsection{Post-development Comparison of the Global-in-Time Duhamel--Picard Method and PINO}
\label{sec:results_comparison}

In this subsection, we compare the post-development resist profiles obtained from the global-in-time Duhamel--Picard method and the PINO surrogate in the previous subsection.

For each method, the protected-group concentration $\prot(\mathbf{x},t_{\mathrm{PEB}})$ is converted into a local dissolution-rate field using the inverse Mack model for negative-tone development~\cite{gao2011inverse_mack}. 
The model parameters used in this test are $a=0.01$, $n=6$, $R_{\max}=100~\mathrm{nm/min}$, and $R_{\min}=0.001~\mathrm{nm/min}$.
The local rate field is subsequently used as the normal propagation speed of the developer front. 
The developer arrival time ${T}(\mathbf{x})$ is determined from the static Eikonal equation using the local dissolution-rate field $R(\mathbf{x})$,
\begin{equation}
	R(\mathbf{x})
	\left|\nabla{T}(\mathbf{x})\right|
	=1,
	\label{eq:development_arrival_time}
\end{equation}
with ${T}=0$ imposed at the top resist surface. 
This arrival-time formulation is commonly used for front propagation and photolithographic development simulations \cite{rouy1992viscosity,sethian1996fastmarching,sethian1996photolithography}.
In the present implementation, the Eikonal equation is solved using a first-order Godunov upwind discretization \cite{rouy1992viscosity,sethian1999traveltime} through synchronous GPU-parallel fixed-point iterations instead of the heap-based fast marching algorithm. 
Periodic boundary conditions are applied laterally, whereas the vertical direction is nonperiodic. 
For this test, the post-development resist profile is obtained by applying a simple threshold cut to the developer arrival-time field ${T}(\mathbf{x})$ at $t_{\mathrm{dev}}=80~\mathrm{s}$.
The resulting three-dimensional resist profiles are then used to quantify the post-development discrepancy between the global-in-time Duhamel--Picard method and the PINO surrogate.

\begin{figure}[H]
    \centering
    \begin{tikzpicture}
        \node[anchor=south west, inner sep=0] (image) at (0,0)
            {\includegraphics[width=1\textwidth]{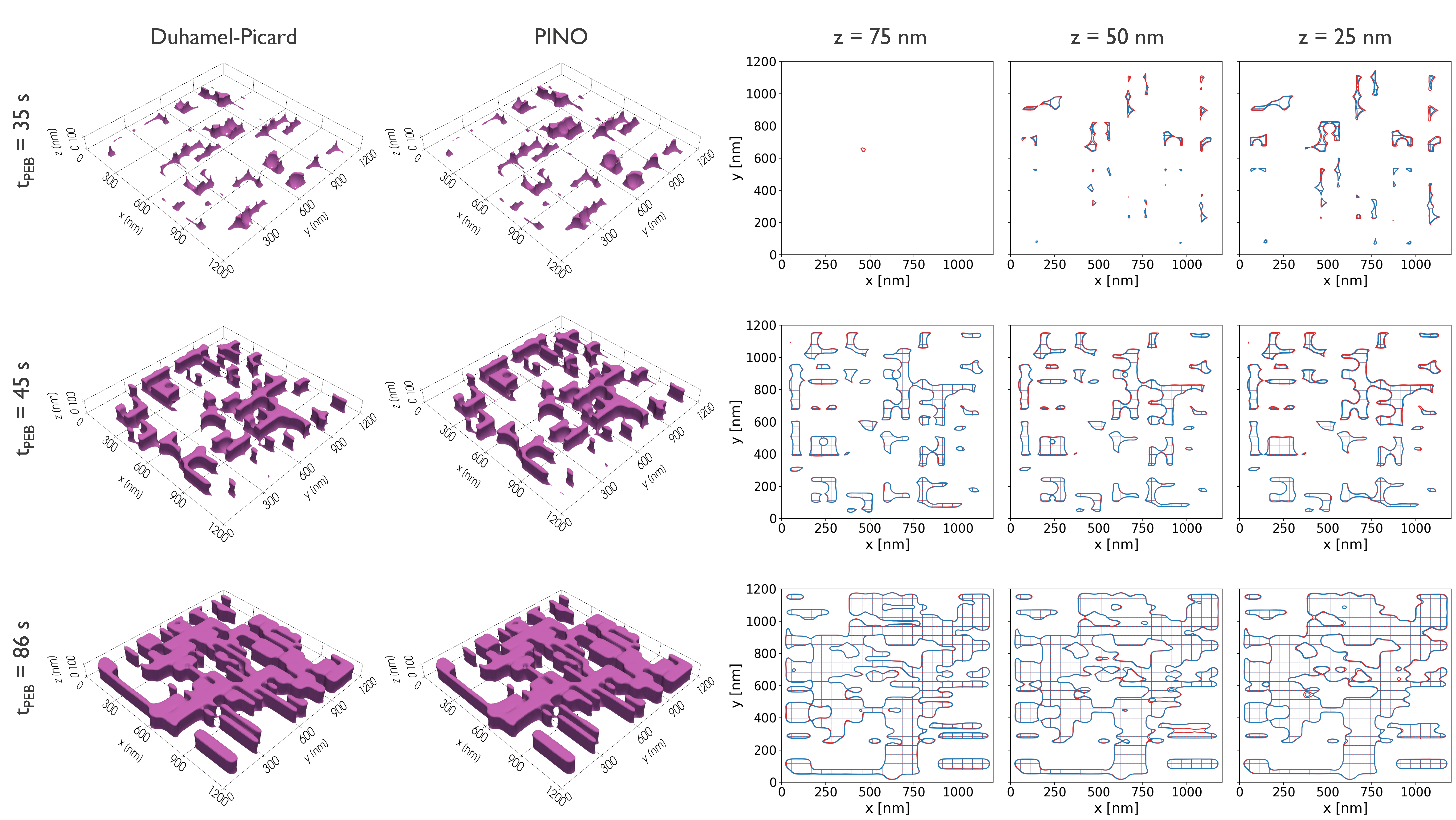}};
        \begin{scope}[
            shift={(image.south west)},
            x={(image.south east)},
            y={(image.north west)}
        ]
            \node[anchor=north west, font=\bfseries, inner sep=1.5pt]
                at (0.02,1.07) {(a)};
            \node[anchor=north west, font=\bfseries, inner sep=1.5pt]
                at (0.48,1.07) {(b)};
        \end{scope}
    \end{tikzpicture}
    \caption{Comparison of post-development resist profiles predicted by the global-in-time Duhamel--Picard method and the PINO surrogate for randomly sampled mask layouts and PEB times. (a) Three-dimensional resist profiles after development at $t_{\mathrm{PEB}}=35$, $45$, and $86~\mathrm{s}$. (b) Corresponding lateral contour comparisons at $z=75$, $50$, and $25~\mathrm{nm}$, with the Duhamel--Picard results shown in blue and the PINO predictions in red.}
    \label{fig:results_development}
\end{figure}

In \Cref{fig:results_development}(a), the three-dimensional resist profiles are compared for both randomly sampled mask layouts and PEB times, while in \Cref{fig:results_development}(b), the corresponding lateral contours are overlaid at different $z$-cuts to provide a rough visual assessment of the discrepancies. 
From the post-development resist profiles, critical dimensions (CDs) are extracted from contour crossings along 40 horizontal and vertical scan lines at different $z$-cuts.

At the measurement plane closer to the resist top surface ($z=75~\mathrm{nm}$), the mean absolute CD error is approximately $2.7~\mathrm{nm}$, with a standard deviation of $2.4~\mathrm{nm}$.
As the measurement plane moves deeper into the resist, the mean absolute CD errors increase to $3.7~\mathrm{nm}$ at $z=50~\mathrm{nm}$ and $5.2~\mathrm{nm}$ at $z=25~\mathrm{nm}$, with corresponding standard
deviations of $3.2~\mathrm{nm}$ and $4.2~\mathrm{nm}$, respectively.
Small differences in $\prot$ between the PINO and Duhamel--Picard predictions are likely mapped nonlinearly to local dissolution-rate differences by the inverse Mack model. 
These differences may then accumulate along the development-front propagation paths, resulting in discrepancies in the developer arrival time.
The final thresholding at ${T}=t_{\mathrm{dev}}$ may further translate small arrival-time differences into finite contour displacements, with this effect likely being more pronounced near narrow, curved, or fragmented resist features.
Consequently, the lower resist sections exhibit broader CD-error distributions even when the underlying $\prot$ fields remain close. 
The extent of this error accumulation is expected to depend on the development model parameters adopted for different resists.

Overall, these results confirm that the PINO largely preserves the principal three-dimensional post-development topology across the randomly sampled mask layouts and PEB times while providing a modest runtime benefit. 
Further study is warranted to address the current memory limitations during training, improve the accuracy of the PINO surrogate, and further reduce the computational runtime.

\section{Conclusion}
\label{sec:conclusion}
In this work, we have presented a global-in-time spectral Duhamel--Picard method for modeling post-exposure bake (PEB) processes.
We have demonstrated that this method accurately and efficiently evaluates the acid, quencher, protected-group, and deprotected-group concentration trajectories over the entire PEB interval.
The polynomial expansions represent these trajectories using a small set of temporal coefficient fields, while Gaussian quadrature enables accurate and efficient evaluation of the temporal integrals.
Using this method, a runtime on the order of one minute was achieved for a $1200\times1200\times100$ computational domain, demonstrating its efficiency for large-scale simulations.

Additionally, we have presented a physics-informed neural operator (PINO) based on the same global-in-time polynomial representation, with the Duhamel formulation used to construct the physics-informed loss.
The PINO predictions show good agreement with the global-in-time Duhamel--Picard results and reliably capture the principal geometric features of the post-development resist profiles.
The three-dimensional profiles predicted by the PINO closely match those obtained using the global-in-time Duhamel--Picard method at $z$-planes closer to the resist surface.
The CD errors increase at deeper $z$-planes, likely because differences in the predicted dissolution-rate fields accumulate over longer development paths.
Nevertheless, the PINO preserves the principal three-dimensional post-development topology across randomly sampled mask layouts and PEB times while reducing the runtime relative to the global-in-time Duhamel--Picard method.
Further research is required to address the current memory limitations during training, improve the accuracy of the PINO surrogate, and further reduce the overall computational cost.

\section{Acknowledgments}
The authors express gratitude to Seungjin Lee for valuable discussions during the initial stages of this work. This work has been enabled in part by the NanoIC pilot line. The acquisition and operation are jointly funded by the Chips Joint Undertaking, through the European Union's Digital Europe (101183266) and Horizon Europe programs (101183277), as well as by the participating states Belgium (Flanders), France, Germany, Finland, Ireland and Romania. For more information, visit nanoic-project.eu.

\bibliography{references}

\end{document}